# Demonstration of a novel dispersive spectral splitting optical element for cost-effective photovoltaic conversion

Carlo Maragliano[1,2], Tim Milakovich[2], Matteo Bronzoni[3], Stefano Rampino[3], Eugene A. Fitzgerald[2], Matteo Chiesa[1] and Marco Stefancich[3]*

[1] Laboratory for Energy and NanoScience (LENS), Center for Future Energy Systems (iFES), Masdar Institute of Science and Technology, P.O. Box 54224, Abu Dhabi, UAE.

[2] Massachusetts Institute of Technology, Material Science and Engineering department, Cambridge, MA 02139 USA

[3] Istituto Materiali per l'Elettronica ed il Magnetismo, Consiglio Nazionale delle Ricerche, Parco Area delle Scienze 37/A - 43124 Parma, Italy

## Abstract

In this letter we report the preliminary validation of a low-cost paradigm for photovoltaic power generation that utilizes a prismatic Fresnel-like lens to simultaneously concentrate and separate sunlight into continuous laterally spaced spectral bands, which are then fed into spectrally matched single-junction photovoltaic cells. A prismatic lens was designed using geometric optics and the dispersive properties of the employed material, and its performance was simulated with a ray-tracing software. After device optimization, it was fabricated by injection molding, suitable for large-scale mass production. We report an average optical transmittance of ~ 90% over the VNIR range with spectral separation in excellent agreement with our simulations. Finally, two prototype systems were tested: one with GaAsP and c-Si photovoltaic devices and one with a pair of copper indium gallium selenide based solar cells. The systems demonstrated an increase in peak electrical power output of 51% and 64% respectively under white light illumination. Given the ease of manufacturability of the proposed device, the reported spectral splitting approach provides a cost-effective alternative to multi-junction solar cells for efficient light-to-electricity conversion ready for mass production.

Photovoltaic cells spectral response depends intrinsically on the nature of the material they are fabricated from. Semiconducting materials absorb photons with energies greater than the band gap and, while photons with energy just above the band gap energy are efficiently converted, those with much larger energies are converted inefficiently due to hot carriers generation and those below band gap are completely lost. Due to the spectral extension of the sunlight (approximately from 350 to 2000 nm, corresponding to a range of energies from 3.55 to 0.62 eV), the maximum theoretical efficiency for a single-junction photovoltaic device under one sun illumination is limited to around 33%[1, 2].

Multi-junction (MJ) solar cells, in which different mono-crystalline materials are vertically stacked on top of each other, alleviate the spectral mismatch issue of single-junction devices and are capable of efficiencies above 40%[3, 4]. However, the stacking of different materials brings several challenges: MJ solar cells require a tight lattice matching, thus imposing a severe constraint on the materials choice, while sub-cells current matching requirement due to the series connection further limit the design freedom. The fabrication process of MJ cells, generally based on mono-crystalline III-V materials, requires expensive epitaxial growth techniques, making the technology economically viable for terrestrial application only under very high concentration (around or above 300x)[4]. Although possible solutions to the drawbacks of MJ cells have been proposed in the literature[5-7], such approach still remains limited to niche applications.

An alternative approach relies on separating the sunlight into laterally spaced wavelength bands, which are then directed to band-matched absorbers. With respect to MJ solar cells, the spectral splitting approach allows overcoming the lattice-matching limit: given that cells are physically independent, each absorber can be independently manufactured with any technique and material. Moreover, current matching among different absorbers is not necessarily required as different converting circuits or parallel/series connections can be devised for each homogeneous cell group. Spectral splitting systems, allowing for cheaper cells, open the way to lower concentration and higher acceptance angle optics compared to MJ solar cell systems, leading to more relaxed requirements on optics and mechanical solar tracking of the setup[8].

Several designs have been proposed in the literature to achieve spectral separation, an extensive review of which can be found in [8] and [9]. While these concepts have proven to be effective for spectral splitting, they are not suitable for low-cost industrial scale manufacturing. In this Letter we describe the preliminary use of a prismatic optical device conceived from the very beginning to fulfill the requirements of mass production. Our optical element is designed to efficiently separate, based on material dispersion, and concentrate incident light into continuous 'rainbow-like' bands that illuminate laterally separated photovoltaic devices of suitable band gaps. Given the low-cost of the optical device and of the power converting cells, our system provides a cost-competitive highly efficient energy generation solution ready for industrial scaling.

The optical element is designed as a set of solid prisms, each splitting a white light beam into its different colors according to material dispersion[10]. The prisms are spatially arranged along a curved surface so that light rays of a specific wavelength converge to the same region at a distance Z (36 cm) from the element, as illustrated in Fig. 1a.

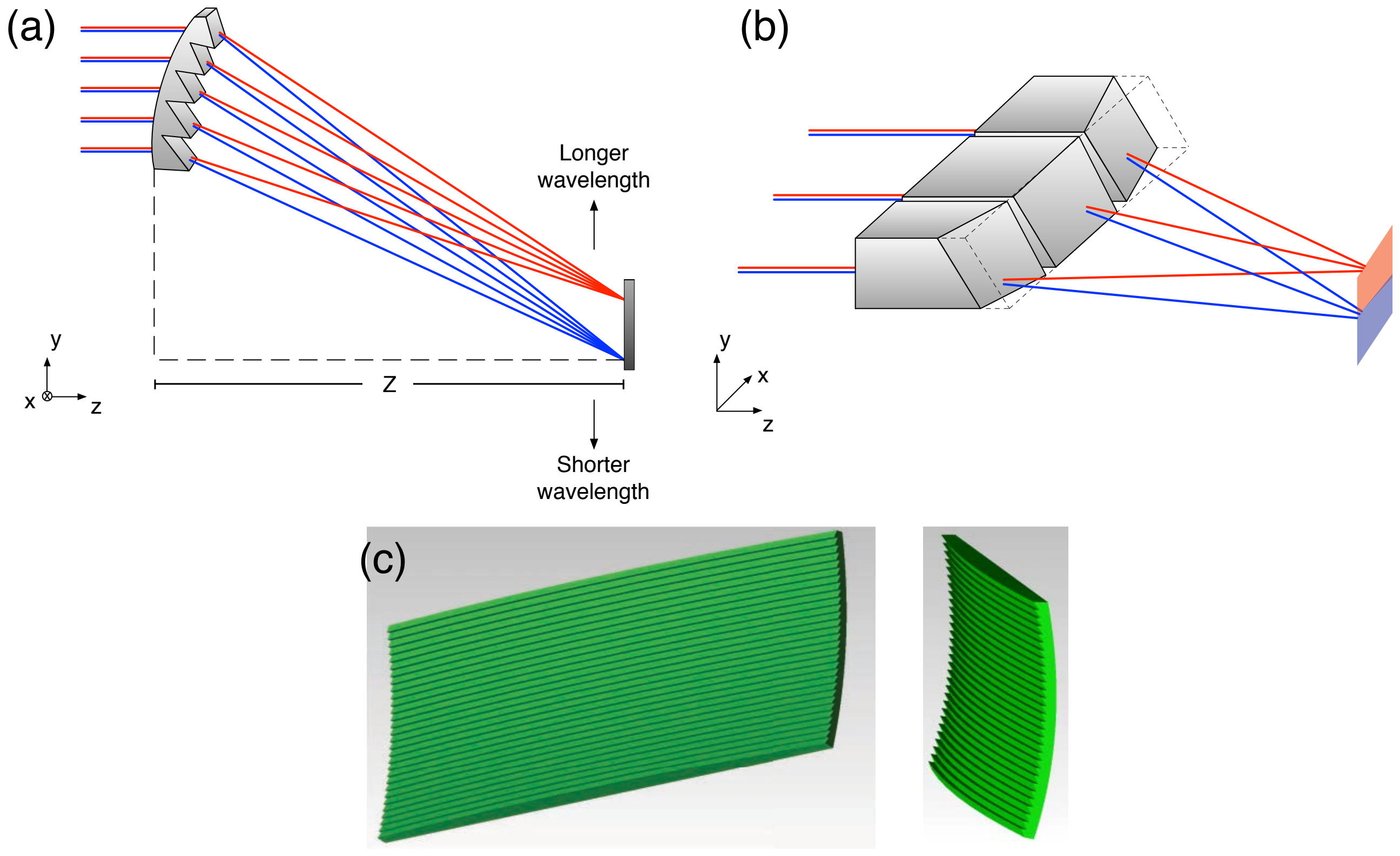

Fig. 1 Conceptual drawing of the optical element: (a) xz section (b) 3D representation. The light rays of a given wavelength converge in the same point at the observation plane, which is situated 36 cm away from the plane of the optical element. Concentration in the x-direction is obtained by properly tilting the exit facet of each section; (c) Design of the point-focus spectral splitter concentrator. The element is composed of 30 trapezoidal prisms arranged in a curved line

The superposition of each prism contribution results in a concentrated and spectrally divided beam along the y direction. To obtain concentration also in the x-direction we conceptually divided each prism into an odd number of sections, as described in Fig. 1b, and each exit facet was properly tilted of a small angle *u* around the vertical axis parallel to y to superimpose the section-generated 'image' with the one of the central portion. The angle *u* for each section was calculated by using Snell's law so to preserve continuity of the element facets. Polycarbonate was chosen as manufacturing material for its transparency in the VNIR region and for its distinctive dispersion curve[11]. Fig. 1c shows the final design of the device. The approximate area of the device is 7x3 $cm^2$, with an average thickness of 2 mm, containing 30 prisms. The lateral size of each prism is 5 mm, which can be further reduced to accomodate more prisms in the same area.

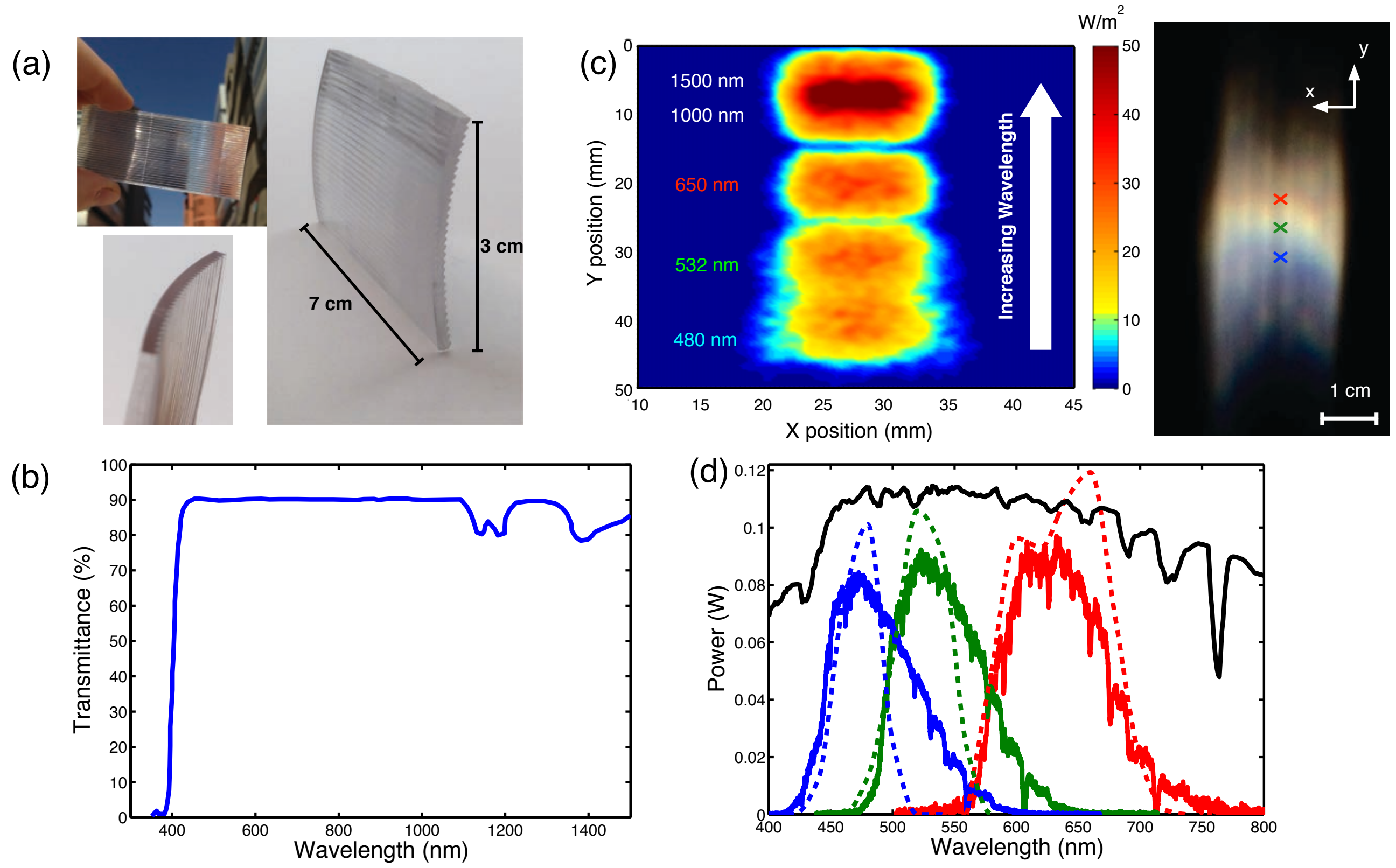

Fig. 2 (a) Pictures of the optical device fabricated by injection molding; (b) Transmittance of the device measured along the range 350-1500 nm; (c) Simulated and experimental light pattern obtained at the observation plane. For the simulations, an intensity of 10 W/$m^2$ was set at the source; (d) Spectral content of the light pattern (solid lines) measured on three different spots along the y direction with an integrating sphere connected to a spectrometer. The separation along the y-direction between contiguous measurements was 0.5 cm. Dashed lines show the simulated spectra.

Fig. 2a shows the optical device fabricated by injection molding, a low-cost technique in which the material is fed into a heated barrel, mixed, and finally forced into a mold cavity. For this technique an initial investment (around 10 k$) is needed to fabricate the mold, which allows for low-cost large-scale manufacturing of the optical device by sample replication. Fig. 2b shows the measured transmittance over the range 400-1500 nm: an average transmittance of ~ 90% in the VNIR range is reported, confirming the good level of transparency of the device. In order to verify and optimize the design, we imported our model into a commercial ray-tracing tool and realistic optical simulations were performed. Figure 2c shows a simulated two-dimensional map of the light intensity obtained at the receiver surface for different wavelengths (480, 532, 650, 1000 and 1500 nm) together with a picture of the experimental light pattern. Note that the simulated map and experimental results are in good agreement. To further validate our results we carried out a spectral analysis with a moving probe feeding to a spectrometer. Fig. 2d shows the light spectra measured using an integrating sphere on three different spots indicated by the markers in Fig. 2c. The figure confirms the spectral separation and the good agreement with simulations, further validating our results.

We performed electrical characterization with both GaAsP/Si cells and a pair of copper indium gallium selenide (CIGS) cells with different band gaps. GaAsP was designed with a band gap of 1.7 eV, while the band gap of the two CIGS cells was 1.16 and 1.3 eV respectively for the low (l-CIGS) and high (h-CIGS) Ga content cells. The two pairs of cells were placed on a x-y-z scanning stage at a distance Z from the optical device and were aligned with the light pattern by maximizing their power output. The cells (dimensions 4x4 mm) were positioned in close proximity so that no or little radiation was lost in between the two. In order to maintain a small divergence angle of the light, the source was positioned at a distance of around 1.5 m from the splitter, which resulted in the incident intensity at the splitter plane being approximately 25 times smaller than that of the standard AM1.5 spectrum [11].

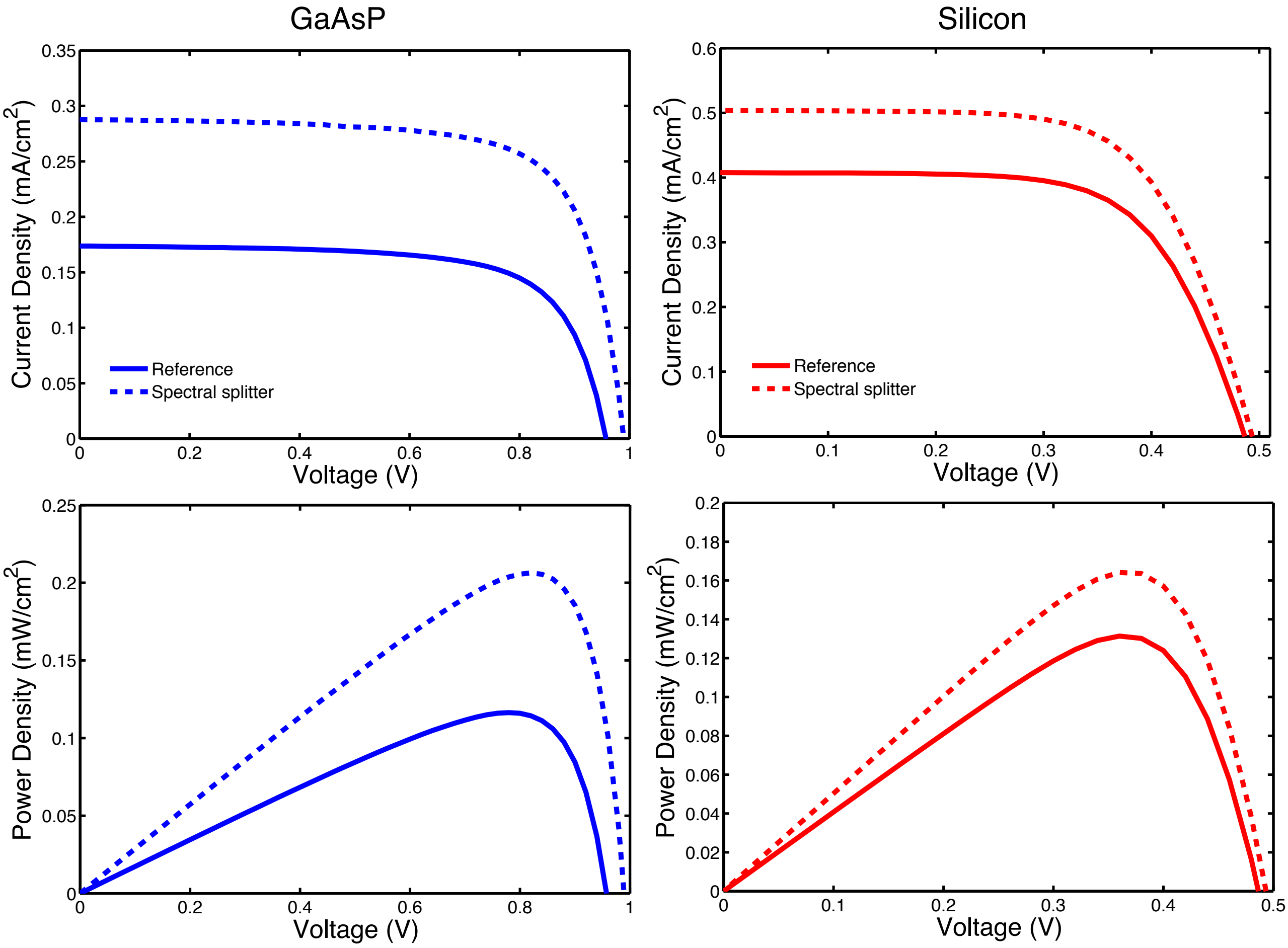

Fig. 3 Current and power density versus voltage curves for GaAsP and silicon cells. Solid lines show reference measurements (full spectrum) while dashed lines refer to those with the concentrator/spectral splitter.

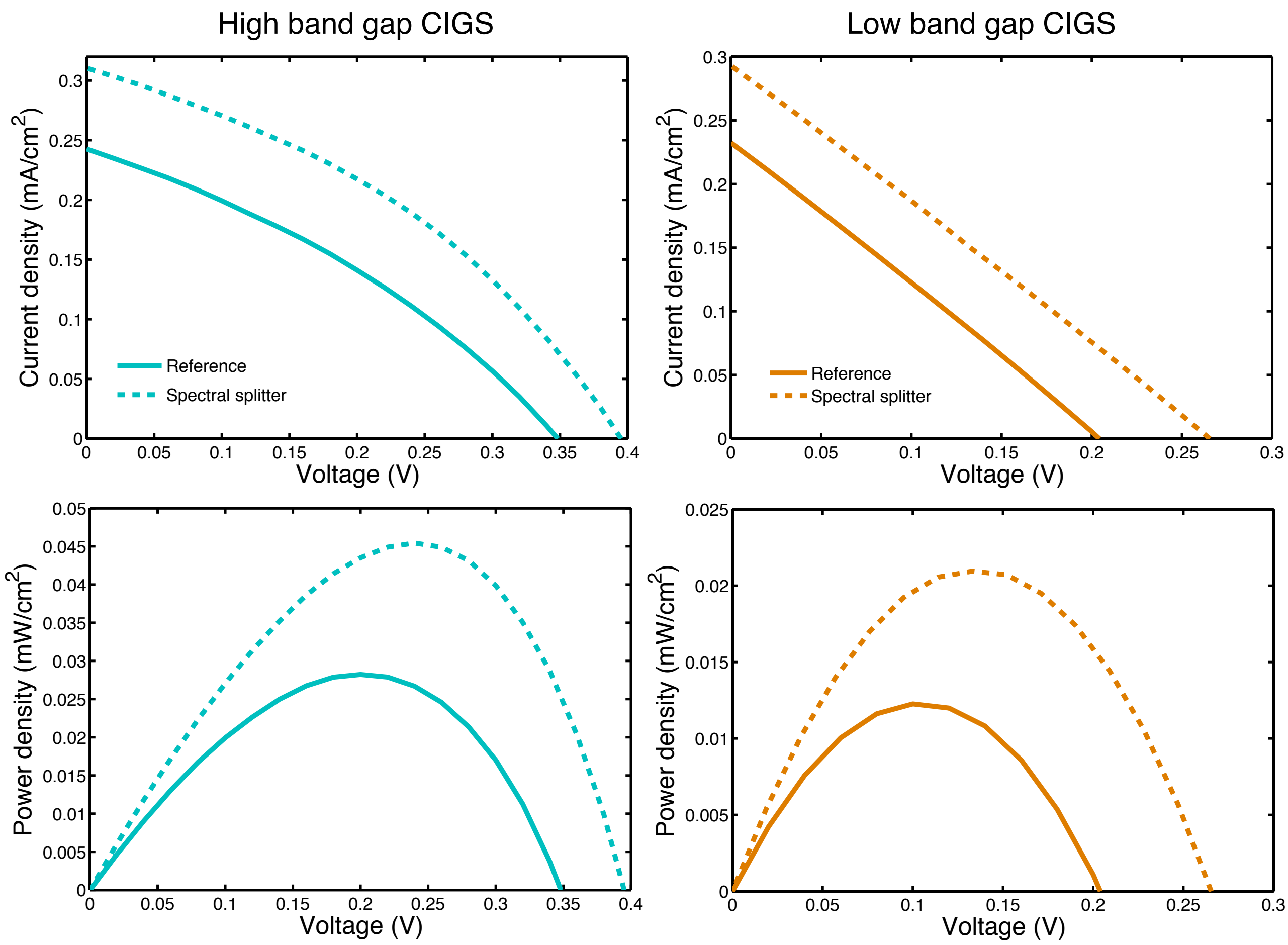

Fig. 4 Current and power density versus voltage curves for CIGS cells. Solid lines show reference measurements (full spectrum) while dashed lines refer to those with the concentrator/spectral splitter.

Figs. 3 and 4 show current and power density versus voltage curves measured on the two couples of cells (GaAsP/Si and the two CIGS cells respectively). Reference (full spectrum) measurements are shown in solid lines, while those with the spectral splitter are shown as dashed lines. The high-band-gap cells (GaAsP and h-CIGS) are illuminated with high-energy photons while the low-energy photons are redirected to the low-band-gap cell (Si and l-CIGS). The incoming light is concentrated with concentration factors in the low-medium range (few to tens x) being function of the wavelength and the divergence angle of the incoming light[11]. For the GaAsP-Si couple we measure optical concentrations of 1.7x and 3x respectively, while for the two CIGS cells we report concentrations of 2x and 4x. Our results demonstrate that the total output power density (added for both cells) is increased with the spectral splitter by 51% and 64% for GaAsP/Si and CIGS respectively. For the CIGS cells we note that the fill factor is low because at low illumination intensities even a small shunt can have a significant effect [12].

Raytracing simulations indicate that under sunlight illumination an increase in output power density of more than 400% is achievable given the small angular divergence of the source (around 4.8 mRad) [11]. As our device generates a continuous spectrally separated light pattern, it is designed to accomodate more than two solar cells, allowing for theoretical conversion efficiencies as high as 38 % for 6 band gaps [11]. Although III–V materials such as GaAsP are currently too expensive to be used under low concentration, spectral splitting allows for a large set of spectrally matched converters, possibly outsourced among different technologies. CIGS seems to be in that sense an interesting candidate as its band gap can be tuned (changing the relative ratio of In and Ga) without significant process modifications[13]. Given the low-cost and potential for large-scale mass production, we believe that the proposed approach provides a cost-effective solution compared to MJ cells for energy generation and thus plays a major role in the wide adoption of spectrum-splitting technology.

We thank Yamila Omar for assistance and fruitful discussions and Harry Apostoleris for proofreading the manuscript. C. M. was supported by Masdar Institute-MIT exchange program.